\documentclass[final,3p,times]{elsarticle}

\usepackage{lineno,hyperref}
\usepackage{amsthm,amsmath,amssymb,eucal}
\usepackage{caption}
\usepackage{subcaption}
\usepackage{mathdots}

\modulolinenumbers[5]

\journal{Physics Letters A}

\newcommand{\ee}{\mathrm{e}}
\newcommand{\N}{\mathbb{N}}
\newcommand{\R}{\mathbb{R}}

\newcommand{\abs}[1]{\left|{#1}\right|}

\newdefinition{remark}{Remark}

\begin{document}

\begin{frontmatter}

\title{Spectral properties of hexagonal lattices with the $-R$ coupling}





\author[label1,label2]
{Pavel Exner}
\ead{exner@ujf.cas.cz}
\author[label2,label3]
{Jan Peka\v{r}}
\ead{honzapekar28@gmail.com}
\address[label1]{Doppler Institute for Mathematical Physics and Applied Mathematics, Czech Technical University,
B\v rehov{\'a} 7, 11519 Prague, Czechia}
\address[label2]{Department of Theoretical Physics, Nuclear Physics Institute, Czech Academy of Sciences, 25068 \v{R}e\v{z} near Prague, Czechia}
\address[label3]{Faculty of Mathematics and Physics, Charles University, V Hole\v{s}ovi\v{c}k\'ach 2, 18040 Prague, Czechia}

\begin{abstract}
We analyze the spectrum of the hexagonal lattice graph with a vertex coupling which manifestly violates the time reversal invariance and at high energies it asymptotically decouples edges at even degree vertices; a comparison is made to the case when such a decoupling occurs at odd degree vertices. We also show that the spectral character does not change if the equilateral elementary cell of the lattice is dilated to have three different edge lengths, except that flat bands are absent if those are incommensurate.
\end{abstract}


\end{frontmatter}

%\linenumbers

%%%%%%%%%%%%%%%%%%%%%%%%%%%%%%%%%%%%%%%
\section{Introduction} %\label{s:intro}

Quantum graphs represent a wide class of systems with a number of interesting properties; for an introduction and rich bibliography we refer to the monographs \cite{BB13, KN22, Ku24}. The versatility of quantum graph models comes, in particular, from the fact that there are numerous ways in which their Hamiltonians can be made self-adjoint, coming from different choices of the ways in which the wave functions are matched at the vertices. Denoting by $\Psi=\{\psi_j\}$ and $\Psi'=\{\psi'_j\}$ the vectors of boundary values of the functions and their derivatives (conventionally taken in the outward direction), respectively, on edges meeting at a vertex of degree $N$, conservation of the probability current at the vertex is guaranteed whenever the condition
 %----------------%
	\begin{equation}\label{Boundary_condition}
		(U-I){\Psi}+i\ell(U+I)\Psi'=0
	\end{equation}
 %----------------%
holds, where $U$ is an $N\times N$ unitary matrix and $\ell$ is a parameter fixing the length scale. This fact is usually referred to the paper \cite{KS99} but the condition was known already to Rofe-Beketov \cite{RB69}.

In the overwhelming part of the quantum graph literature this richness lies fallow, though, as most often people use the so-called $\delta$ coupling where the wave functions are continuous at the vertex (having a common value $\psi$) and $\sum_j \psi'_j=\alpha\psi$ with some $\alpha\in\R$, especially the particular choice $\alpha=0$ referred to as Kirchhoff. There are, however, other choices, some of the potential physical interest. Motivated by attempts to use quantum graphs to model the anomalous Hall effect \cite{SK15, SV23}, we proposed in \cite{ET18} a simple vertex coupling violating the time-reversal invariance, the violation `intensity' being maximal at the momentum $k=l^{-1}$; the corresponding unitary matrix is $R$ given by \eqref{Intr_-R_matrix} below. One of the interesting properties of this coupling is that its high-energy transport properties depend on the vertex parity: if $N$ is odd, the edges become asymptotically decoupled while for even-degree vertices the scattering matrix has a nontrivial limit. One of the consequences concerns the ratio of gap-to-band size in periodic graphs with such a coupling -- see, e.g., \cite{BE22, BET22, ET18}. Note that the $R$ coupling belongs to a wider class in which the matrices $U$ are \emph{circulant} and \emph{transpose non-invariant}; such couplings exhibit, for instance, a non-trivial $\mathcal{PT}$-symmetry even if the corresponding Hamiltonians are self-adjoint \cite{ET21}.

The true reason behind the mentioned high-energy dichotomy is not that much the parity itself, but rather the spectrum of the matrix $U$. The asymptotic edge decoupling occurs whenever $-1$ is \emph{not} an eigenvalue of $U$; recall that the spectrum of $R$ consists of the complex roots of unity. We have illustrated this fact in \cite{ET21} on the example of a periodic square lattice graph with the vertex coupling given by $U=\ee^{i\mu}R$: unless $\mu$ was an integer multiple of $\frac{2\pi}{N}$ the gaps in the spectrum expanded so that asymptotically they dominated the spectrum. The main aim of the present letter is to show that the said dichotomy can be \emph{reverted}, making the odd degree vertices `transport friendly' at high energies. To this aim is enough to choose $\mu=\pi$, in other words, to consider the coupling described by the matrix
 %----------------%
\begin{equation}\label{Intr_-R_matrix}
		U := -R = \begin{pmatrix} 0 & -1 & 0 & {\dots} & 0 \\
			0 & 0 & -1 & {\dots} & 0 \\
			{\vdots} & 0 & 0 & {\ddots} & {\vdots} \\
			0 & {\dots} & {\ddots} & {\ddots} & -1 \\
			-1 & 0 & {\dots} & 0 & 0 \\
		\end{pmatrix},
\end{equation}
 %----------------%
which in the component form reduces to
 %----------------%
\begin{equation}\label{Boundary_component}
    -\psi_{j+1} - \psi_j + i\ell(-\psi'_{j+1} + \psi'_j) = 0, \; j \in \mathbb{Z}\; (\text{mod}\;N).
\end{equation}
 %----------------%
We will first analyze the spectral and scattering properties of a single vertex. After that we will investigate a regular honeycomb lattice on which the band-to-gap ratio tends to zero as $k\to\infty$ in case of the $R$ coupling \cite{ET18}; we will show that for the coupling \eqref{Boundary_component} the opposite is true. Finally, we will look at the spectrum of deformed periodic honeycomb lattices with the $-R$ coupling in the spirit of \cite{ET15}; we will see that while the band dominance is preserved, geometric deformation may lead to more than one asymptotic behavior type of the gaps.

%%%%%%%%%%%%%%%%%%%%%%%%%%%%%%%%%%%%%
\section{Star graphs} %\label{s:star}

Consider a star graph with $N$ semi-infinite edges meeting at a single vertex and the vertex condition \eqref{Boundary_component}. By general principles \cite[Cor.~1 to Thm.~8.19]{We80}, the negative spectrum of such a system is discrete. It can be found easily: using the Ansatz ${\psi_j} = {c_j}\,\ee^{-{\kappa}x}$, $\kappa > 0$, the requirement \eqref{Boundary_component} yields a system of equations for the coefficients $c_j$ which turns up to be solvable \emph{iff}
 %----------------%
\begin{equation}
    (-1-i{\kappa}\ell)^{N} + (-1)^{N-1}(-1+i{\kappa}\ell)^{N} = 0.
\end{equation}
 %----------------%
This condition has solutions only if $N \ge 3$, and the eigenvalues of our Hamiltonian are $-{\kappa}^2$, where
 %----------------%
\begin{equation}\label{star_bound_states}
    \kappa = \frac{1}{\ell\tan{\frac{m\pi}{N}}}
\end{equation}
 %----------------%
with $m$ running through $1,\:{\dots}\:,\: {\big\lfloor}\frac{N}{2}{\big\rfloor}$ for $N$ odd and $1,\:{\dots}\:,\: {\big\lfloor}\frac{N-1}{2}{\big\rfloor}$ for $N$ even, in particular, for $N=3,4$ there is a single negative eigenvalue equal to $\frac13 \ell^{-2}$ and $\ell^{-2}$, respectively.

As for the continuous spectrum, the on-shell S-matrix at momentum $k$ is by \cite[Sec.~2.1.1]{BK13} equal to
 %----------------%
\begin{equation} \label{smatrix}
    S(k) = \frac{(k\ell-1)I+(k\ell+1)U}{(k\ell+1)I+(k\ell-1)U}.
\end{equation}
 %----------------%
As indicated in the introduction, its behavior as $k\to\infty$, and likewise as $k\to 0$, is determined by the spectrum of the matrix $U$; it is obvious from \eqref{smatrix} that the limits are trivial as long as $-1$ and $1$, respectively, do \emph{not} belong to $\sigma(U)$. In our current situation with $U$ given by \eqref{Intr_-R_matrix}, its spectrum consists of eigenvalues $-\omega_j= -\ee^{2{\pi}ij/N}$, $\,j = 0,\,\dots,\,N-1$. Hence $-1 \in {\sigma}(U)$ holds always; a direct computation using the fact that both $I$ and $U$ are circulant matrices gives
 %----------------%
\begin{equation} \label{smatrix_inf}
    \lim_{k \to \infty} S_{ij}(k) = \frac{N-2}{N}\,\delta_{ij} -\frac{2}{N}\,(1-\delta_{ij}).
\end{equation}
 %----------------%
In a similar way, we get
 %----------------%
\begin{equation} \label{smatrix_0}
    \lim_{k \to 0} S_{ij}(k) = \frac{2-N}{N}\,\delta_{ij} + (-1)^{i-j}\,\frac{2}{N}\,(1-\delta_{ij})
\end{equation}
 %----------------%
for an even $N$, otherwise $\lim_{k \to 0} S(k) = -I$.

%%%%%%%%%%%%%%%%%%%%%%%%%%%%%%%%%%%%%%%%%%%%%%
\section{Hexagonal lattice} %\label{s:hexagon}

\begin{figure}[b]\centering
		\includegraphics[trim = {2cm 15cm 10cm 6cm},clip]{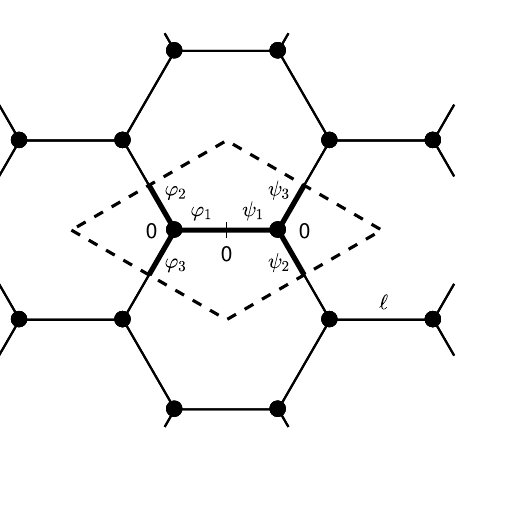}
		\caption{Periodic hexagonal lattice; the elementary cell is highlighted.}
		\label{Hexagon_periodic}
	\end{figure}

\begin{figure}[b]\centering
		\includegraphics{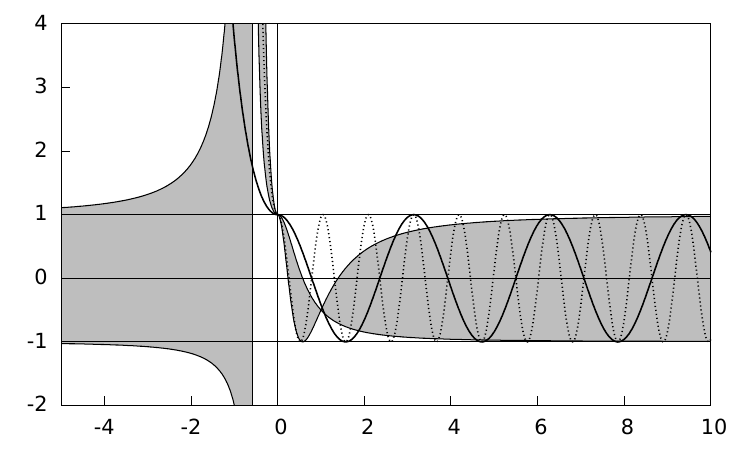}
		\caption{Spectral condition solution for a hexagonal lattice with the $-R$ vertex condition; the full line corresponds to $l = 1$, while the dashed one to $l = 3$. The intersection referring to the second negative band for $l = 3$ is outside the depicted area.}
		\label{Hexagon_solution}
	\end{figure}

Consider next a periodic hexagonal lattice with the edges of length $l$ sketched in the Fig.~\ref{Hexagon_periodic}.
 %----------------%
 %----------------%
Since the latter fixes the length scale, without loss of generality we can put $\ell=1$ in \eqref{Boundary_component}; the coordinate direction is chosen to be increasing from the left to the right. As usual in the case of periodic graphs \cite[Sec.~4.2]{BK13}, one can use the Bloch-Floquet decomposition and reduce the problem to finding the spectrum on the period cell. To this aim, we use the Ansatz
 %----------------%
	\begin{equation}\label{Hex_Ans}
		\begin{split}
			{\psi}_j(x)&= C_j^+\ee^{ikx} + C_j^-\ee^{-ikx},\;\; x \in [0, \textstyle{\frac{l}{2}}],\\
			{\varphi}_j(x)&= D_j^+\ee^{ikx} + D_j^-\ee^{-ikx},\;\; x \in [-\textstyle{\frac{l}{2}},0],\\
		\end{split}
	\end{equation}
  %----------------%
with $j = 1,\:2,\:3$. At the cell center, functions $\psi_1$ and $\varphi_1$ have to be matched smoothly which yields
    %----------------%	
   \begin{equation}\label{Hex_D_1}
	D^+_1 = C^+_1,\;\; D^-_1 = C^-_1,
	\end{equation}
   %----------------%
while imposing the quasiperiodic conditions at the border of the fundamental domain we get
   %----------------%
  \begin{equation}\label{Hex_D_2_3}
		\begin{split}
			D^+_2 &= C^+_2\ee^{ikl}\ee^{-\theta_2},\; D^-_2 = C^-_2\ee^{-ikl}\ee^{-\theta_2},\\
			D^+_3 &= C^+_3\ee^{ikl}\ee^{-\theta_1},\; D^-_3 = C^-_3\ee^{-ikl}\ee^{-\theta_1},\\
		\end{split}
\end{equation}
     %----------------%
where $\theta_1$ and $\theta_2$ are the quasimomentum components running both through the interval $(-\pi,\pi]$ as the Brillouin zone is the square $(-\pi,\pi]^2$. At the two vertices in the period cell condition \eqref{Boundary_component} must be valid giving
 %----------------%
	\begin{equation}\label{Hex_ver_cond}
		\begin{aligned}
			  -{\psi_{2}(0)} - {\psi_1(l/2)} - i({\psi^{'}_{2}(0)} + {\psi^{'}_{1}(l/2)}) &= 0,\\
                -{\psi_{3}(0)} - {\psi_2(0)} + i(-{\psi^{'}_{3}(0)} +{\psi^{'}_{2}(0)}) &= 0,\\
                - {\psi_1(l/2)} -{\psi_{3}(0)} + i({\psi^{'}_{1}(l/2)}+{\psi^{'}_{3}(0)}) &= 0,\\
                 -{\varphi_{2}(0)} - {\varphi_1(-l/2)} + i({\varphi^{'}_{2}(0)} + {\varphi^{'}_{1}(-l/2)}) &= 0,\\
                -{\varphi_{3}(0)} - {\varphi_2(0)} + i({\varphi^{'}_{3}(0)} -{\varphi^{'}_{2}(0)}) &= 0,\\
                - {\varphi_1(-l/2)} -{\varphi_{3}(0)} - i({\varphi^{'}_{1}(-l/2)}+{\varphi^{'}_{3}(0)}) &= 0.\\
		\end{aligned}
	\end{equation}
  %----------------%
Combining the condition \eqref{Hex_Ans}--\eqref{Hex_ver_cond} we arrive at a system of linear equations for the coefficients $C^{\pm}_j$, which is solvable only if its determinant vanishes; excluding numerical prefactors we obtain the following \emph{spectral condition}:
  %----------------%	
   \begin{equation}\label{Hex_spectral_cond}
	 \sin{kl}\;(\cos{2kl}\;(3k^2 +1)^2 + 3k^4 + 6k^2 -1 -4d_{\theta}k^2(k^2-1)) = 0
	\end{equation}
   %----------------%
with $d_{\theta} := \cos(\theta_1 - \theta_2) + \cos{\theta_1} + \cos{\theta_2}$; $\:d_{\theta} \in [-\frac{3}{2},3]$. Its solutions are of two kinds; we have either $\sin{kl} = 0$ giving rise to \emph{flat bands} at the energies $k^2=\big(\frac{m{\pi} }{l}\big)^2,\: m\in\N$, or those satisfying the relation
  %----------------%	
   \begin{equation}\label{Hex_spectral_cond_pos}
	\cos{2kl} =\frac{1-6k^2-3k^4 + 4d_{\theta}k^2(k^2-1)}{(3k^2 +1)^2}
\end{equation}
   %----------------%
for some $\theta\in(-\pi,\pi]^2$. This applies to the positive part of the spectrum; in the negative part the flat bands are absent and the counterpart to \eqref{Hex_spectral_cond_pos} is obtained simply by substitution $k = i\kappa$, $\:\kappa > 0$, giving
  %----------------%	
   \begin{equation}\label{Hex_spectral_cond_neg}
	 \cosh{2{\kappa}l} =\frac{1+6{\kappa}^2-3{\kappa}^4 + 4d_{\theta}{\kappa}^2({\kappa}^2+1)}{(3{\kappa}^2 -1)^2}.
	\end{equation}
   %----------------%
Both conditions \eqref{Hex_spectral_cond_pos} and \eqref{Hex_spectral_cond_neg} determining the absolutely continuous spectral bands allows for a graphical solution as sketched in Fig.~\ref{Hexagon_solution},
 %----------------%
 %----------------%
where the positive horizontal half-line corresponds to the momentum variable $k$, the negative one to~$\kappa$. It is not difficult to derive the band properties. We begin with the negative part of the spectrum:
 %----------------%
        \begin{itemize}
        \item bands are determined by the intersection of $\cosh{2{\kappa}l}$ with the region bordered by the curves $g_+(\kappa) = \frac{1+18{\kappa}^2 + 9{\kappa}^4}{(3{\kappa}^2-1)^2}$ and $g_-(\kappa) = \frac{1 +3{\kappa}^2}{1-3{\kappa}^2}$,
        \item negative spectrum is never empty, $\inf \sigma(H) < -\frac{1}{3}$,
        \item for $l > 2\sqrt{3}$ the negative spectral bands are strictly negative and there are two of them, one below and one above the energy $-\frac{1}{3}$,
        \item for $2\sqrt{3} \ge l > \sqrt{3}$ the second negative band extends to zero,
        \item for $l \le \sqrt{3}$ there is only one negative band, placed below the energy $-\frac{1}{3}$,
        \item for large $l$ the negative bands become exponentially narrow, centered around the single vertex bound state energy \eqref{star_bound_states}, in this case $-\frac{1}{3}$. They are of the size $\approx \frac{2}{\sqrt{3}}\ee^{-\frac{1}{\sqrt{3}}l}$ up to an $\mathcal{O}\big(\ee^{-\frac{2}{\sqrt{3}}l}\big)$ error, with the distance $\approx \frac{2}{\sqrt{3}}\ee^{-\frac{2}{\sqrt{3}}l} + \mathcal{O}\big(\ee^{-\frac{4}{\sqrt{3}}l}\big)$ separating them, both at the momentum scale,
        \item the first band decreases in the energy scale as $l \to 0$, being between energies $\big(-\frac{2}{\sqrt{3}}\frac{1}{l},-\frac{1}{3}\big)$ up to an $\mathcal{O}(l)$ error.
    \end{itemize}
  %----------------%
 On the other hand, the positive spectral bands are determined by the intersection of $\cos{2{k}l}$ with the region bordered by the curves $h_+(k) = \frac{1-18{k}^2 + 9{k}^4}{(3{k}^2+1)^2}$ and $h_-(k) = \frac{1 -3{k}^2}{1+3{k}^2}$ and one can easily find their properties:
  %----------------%
        \begin{itemize}
        \item the number of gaps in the positive spectrum is infinite; they are centered around the points $k = \frac{m{\pi}}{2l}$. If $m$ is even, the corresponding gap has the asymptotic width ${\Delta}E = \frac{8}{\sqrt{3}}\frac{1}{l} + \mathcal{O}(m^{-2})$ at the energy scale, while for $m$ odd it is ${\Delta}E = \frac{4}{\sqrt{3}}\frac{1}{l} + \mathcal{O}(m^{-2})$ as $m \to \infty$,
        \item the first positive band starts at zero if $\sqrt{3} \le l < 2\sqrt{3}$, otherwise there is a gap between it and the second (or the only) negative band.
        \item at higher energies the spectrum is dominated by spectral bands. Because $\lim\limits_{k \to \infty}h_+(k) = 1$ and $\lim\limits_{k \to \infty}h_-(k) = -1$, the probability of belonging to the positive spectrum in the spirit of \cite{BB13} can be trivially expressed analytically and equals to
        %----------------%
         \begin{equation} \label{Probability_in_positive}
            P_{\sigma}(H) := \lim_{E \to \infty} \frac{1}{E}\abs{\sigma(H) \cap [0,E]} = 1.
        \end{equation}
         %----------------%
    \end{itemize}
 %----------------%
 \begin{remark}
For comparison we show in Fig.~\ref{Hexagon_R_solution}
  %----------------%
	\begin{figure}[b]\centering
		\includegraphics{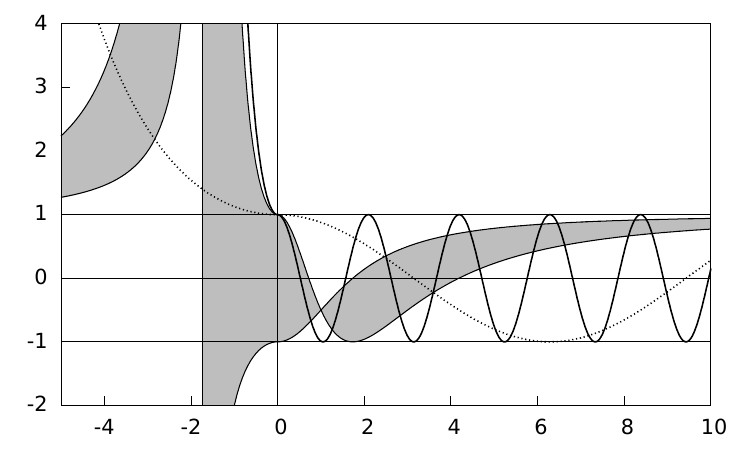}
		\caption{Spectral condition solution for a hexagonal lattice with the $R$ vertex condition; the full line corresponds to $l = \frac{3}{2}$, the dashed one to $l = \frac{1}{4}$.}
		\label{Hexagon_R_solution}
	\end{figure}
 %----------------%
the graphical solution of the spectral problem for hexagonal lattice with the $R$ coupling, correcting at the same time an error in \cite{ET18} concerning the lower boundary of the shaded region; it does not affect the conclusions about the numbers and distribution of bands and gaps, as well as the dominance of the latter, but it changes the coefficient values in the asymptotic expressions. The upper boundary is $g_+(\kappa) = \frac{{\kappa}^4+18{\kappa}^2 + 9}{({\kappa}^2-3)^2}$ and $h_+(k) = \frac{k^4-18{k}^2 + 9}{({k}^2+3)^2}$ in the negative and positive part respectively, while for the lower one the expressions given in \cite{ET18} have to be replaced by $g_-(\kappa) = \frac{{\kappa}^2+3}{{\kappa}^2-3}$ and $h_-(k) = \frac{{k}^2+3}{{k}^2-3}$, respectively. Consequently,
 %----------------%
\begin{itemize}
    \item the two negative bands centered around $-3$ have for large $l$ the widths $\approx 2\sqrt{3}e^{-\sqrt{3}l}$ up to an $\mathcal{O}(e^{-2{\sqrt{3}}l})$ error, with the distance $\approx 2{\sqrt{3}}e^{-2\sqrt{3}l} + \mathcal{O}(e^{-4{\sqrt{3}}l})$ separating them,
     \item the lowest band decreases in the energy scale as $l \to 0$, being contained in $(-\frac{2\sqrt{3}}{l},-\frac{\sqrt{3}}{l})$, up to an $\mathcal{O}(l)$ error,
    \item the first positive band starts at zero if $l \ge \frac{2}{\sqrt{3}}$, otherwise there is a gap between it and the second negative band,
    \item as for the asymptotic behavior of the positive bands, the two around the points $k^2 = (\frac{m\pi}{l})^2$ have the width $\frac{2\sqrt{3}}{l} + \mathcal{O}(m^{-2})$, and the gap between them is $\approx \frac{4\sqrt{3}}{l} + \mathcal{O}(m^{-2})$ as $m \to \infty$.
\end{itemize}
  %----------------%
 \end{remark}
 %----------------%

%%%%%%%%%%%%%%%%%%%%%%%%%%%%%%%%%%%%%%%%%%%%%%%%%%%%%%%%%%
\section{General hexagonal lattice} %\label{s:deformation}

\begin{figure}[b]\centering
		\includegraphics{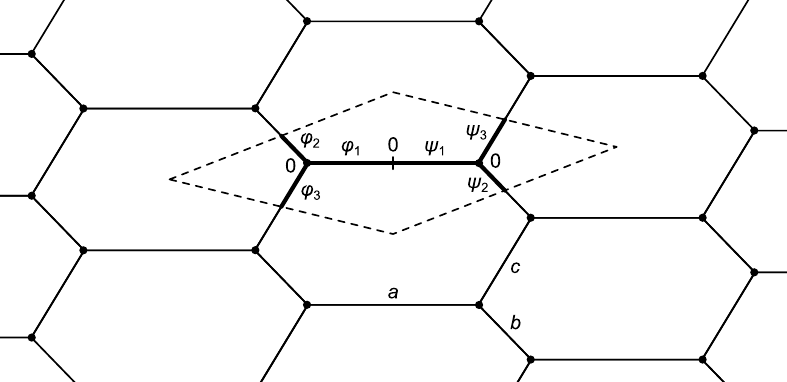} %[trim = {2cm 15cm 10cm 6cm},clip]
		\caption{General periodic hexagonal lattice}
		\label{General_Hexagon}
	\end{figure}

Next we consider the same infinite hexagonal lattice, but now scaled independently in each of the three directions, as pictured in Fig~\ref{General_Hexagon};
 %----------------%
 %----------------%
the Hamiltonian (=Laplacian) on the edges and the vertex coupling $-R$ remain unchanged. To solve the spectral problem we employ an Ansatz analogous to \eqref{Hex_Ans} used in the above particular case,
 %----------------%
	\begin{equation}\label{Dik_hex_Ans}
		\begin{split}
			{\psi}_1(x)&= C_1^+\ee^{ikx} + C_1^-\ee^{-ikx},\;\; x \in [0, \textstyle{\frac{a}{2}}],\\
                {\psi}_2(x)&= C_2^+\ee^{ikx} + C_2^-\ee^{-ikx},\;\; x \in [0, \textstyle{\frac{b}{2}}],\\
                {\psi}_3(x)&= C_3^+\ee^{ikx} + C_3^-\ee^{-ikx},\;\; x \in [0, \textstyle{\frac{c}{2}}],\\
			{\varphi}_1(x)&= D_1^+\ee^{ikx} + D_1^-\ee^{-ikx},\;\; x \in [-\textstyle{\frac{a}{2}},0],\\
                {\varphi}_2(x)&= D_2^+\ee^{ikx} + D_2^-\ee^{-ikx},\;\; x \in [-\textstyle{\frac{b}{2}},0],\\
                {\varphi}_3(x)&= D_3^+\ee^{ikx} + D_3^-\ee^{-ikx},\;\; x \in [-\textstyle{\frac{c}{2}},0].\\
		\end{split}
	\end{equation}
  %----------------%
The smooth matching of the functions along the edge of length $a$ leads again to the condition \eqref{Hex_D_1}, while the Bloch-Floquet conditions now give
 %----------------%
 \begin{equation}\label{Dil_hex_D_2_3}
		\begin{split}
			D^+_2 &= C^+_2\ee^{ikb}\ee^{-\theta_2},\; D^-_2 = C^-_2\ee^{-ikb}\ee^{-\theta_2},\\
			D^+_3 &= C^+_3\ee^{ikc}\ee^{-\theta_1},\; D^-_3 = C^-_3\ee^{-ikc}\ee^{-\theta_1}.\\
		\end{split}
\end{equation}
 %----------------%
The matching relations coming from condition \eqref{Boundary_component} have a structure analogous to \eqref{Hex_ver_cond}, but with the coordinates adjusted to the present edge lengths. The resulting linear system of equations for the coefficients $C^{\pm}_j$ is solvable provided
 %----------------%
 \begin{equation}\label{Dil_hex_spec_cond}
 \begin{aligned}
	&(3k^4+1)\sin{ak}\sin{bk}\sin{ck}\\
 + &2k^2(k^2-1)(\sin{bk}\cos{\theta_2}+\sin{ck}\cos{\theta_1}+\sin{ak}\cos{(\theta_2-\theta_1)})\\
 -&2k^2(k^2+1)(\cos{ak}\cos{bk}\sin{ck}+\cos{bk}\cos{ck}\sin{ak}+\cos{ck}\cos{ak}\sin{bk}) = 0.
 \end{aligned}
\end{equation}
 %----------------%
 Using simple algebraic manipulations, one can check that by choosing $a = b = c = l$ the condition \eqref{Dil_hex_spec_cond} reduces to \eqref{Hex_spectral_cond}.

 Before discussing this model, let us recall the paper \cite{ET15} in which such a general hexagonal lattice was investigated in the situation when the vertex coupling is of a $\delta$-type. Many of the proofs there can be used directly, modulo minor modifications, and we just summarize the conclusions for our present model.

 As for the \emph{flat bands}, similarly to Proposition 2.1, 2.2 and 3.1 of \cite{ET15}, the point spectrum is non-empty only if the edge lengths are rationally related, as in that case there is an infinite number of $k$'s for which $\sin{ak} = \sin{bk} = \sin{ck} = 0$ holds; their specific values, i.e. the flat band's momenta, obviously depend on the ratios involved. For incommensurate lengths, the spectrum is purely absolutely continuous.

 Let us turn to the (non-flat) spectral bands. The condition \eqref{Dil_hex_spec_cond} can be rewritten as
  %----------------%
 \begin{equation}\label{Dil_hex_spec_cond_better}
 \begin{aligned}
&(3k^4+1)+2k^2(k^2-1)\bigg{(}\frac{\cos{\theta_2}}{\sin{ak}\sin{ck}}+\frac{\cos{\theta_1}}
{\sin{ak}\sin{bk}}+\frac{\cos{(\theta_2-\theta_1)}}{\sin{bk}\sin{ck}}\bigg{)}\\
 -&2k^2(k^2+1)(\cot{ak}\cot{bk}+\cot{bk}\cot{ck}+\cot{ck}\cot{ak}) = 0,
 \end{aligned}
\end{equation}
 %----------------%
 provided we exclude the `Dirichlet points', which is a common nickname for the values of $k$ such that $\sin{lk} = 0$ for any of $l \in \{a,b,c\}$. The conclusions we can draw for this condition are less specific than in the regular case; we focus on three particular energy regions.

 %%%%%%%%%%%%%%%%%%%%%%%%%%%%%%%%%%%%%%%%
 \subsection{The high-energy asymptotics}

 For large momentum values, \eqref{Dil_hex_spec_cond_better} can be written as
 %----------------%
  \begin{equation}\label{Dil_hex_spec_cond_high_k}
	3+ 2\bigg{(}\frac{\cos{\theta_2}}{\sin{ak}\sin{ck}}+\frac{\cos{\theta_1}}
{\sin{ak}\sin{bk}}+\frac{\cos{(\theta_2-\theta_1)}}{\sin{bk}\sin{ck}}\bigg{)} -2(\cot{ak}\cot{bk}+\cot{bk}\cot{ck}+\cot{ck}\cot{ak}) = \mathcal{O}(k^{-2}),
\end{equation}
 %----------------%
 or alternatively
  %----------------%
  \begin{equation*}\label{Dil_hex_spec_cond_high_k_uprav}
  \begin{aligned}
	-&(\cot{ak}+\cot{bk}+\cot{ck})^2+\frac{1}{\sin^2{ak}} +\frac{1}{\sin^2{bk}} +\frac{1}{\sin^2{ck}}\\
 +&2\bigg{(}\frac{\cos{\theta_2}}{\sin{ak}\sin{ck}}+\frac{\cos{\theta_1}}
{\sin{ak}\sin{bk}}+\frac{\cos{(\theta_2-\theta_1)}}{\sin{bk}\sin{ck}}\bigg{)}  = \mathcal{O}(k^{-2}).
\end{aligned}
\end{equation*}
 %----------------%
 The left-hand side of this condition coincides with the one derived in \cite{ET15} for the Kirchhoff coupling, where the continuous spectrum covers the whole positive half-line. In our case, the right-hand side does not vanish and gaps are generally present, however, in the language of probability \eqref{Probability_in_positive} we have
 %----------------%
 \begin{itemize}
     \item $P_{\sigma}(H) = 1$ for any choice of the lengths $a,b,c$.
 \end{itemize}
 %----------------%

 Let us now return to the form \eqref{Dil_hex_spec_cond} of the spectral condition and consider the momentum value $k = \frac{m{\pi}}{a}$, $m \in \mathbb{N}$ assuming that all the hexagon lengths are incommensurate. Then the spectral condition reduces to
  %----------------%
 \begin{equation*}
 \begin{aligned}
 + &2k^2(k^2-1)(\sin{bk}\cos{\theta_2}+\sin{ck}\cos{\theta_1})\\
 -(-1)^m&2k^2(k^2+1)(\sin{bk}\cos{ck}+\sin{ck}\cos{bk}) = 0,
 \end{aligned}
\end{equation*}
 %----------------%
 which can be rewritten as
   %----------------%
 \begin{equation*}
 \begin{aligned}
  &k^4[\sin{bk}(\cos{\theta_2}-(-1)^m\cos{ck})+\sin{ck}(\cos{\theta_1}-(-1)^m\cos{bk})]\\
 -&k^2[\sin{bk}(\cos{\theta_2}+(-1)^m\cos{ck})+\sin{ck}(\cos{\theta_1}+(-1)^m\cos{bk})] = 0.
 \end{aligned}
\end{equation*}
 %----------------%}
 It is obvious that the chosen points $k$ cannot belong to the spectrum, as there are no values of $\theta_1$ and $\theta_2$ which would annulate the terms proportional to $k^4$ and to $k^2$ simultaneously. The spectrum is a closed set, hence one has to investigate a neighborhood of such a point. To this aim, we expand the condition \eqref{Dil_hex_spec_cond} for the momentum $\frac{m{\pi}}{a}+\delta$ to the second order, $0 = C+{\delta}B+{\delta}^2A +\mathcal{O}({\delta}^3)$, where
 %----------------%
  \begin{equation*}
 \begin{aligned}
  C=&-2(-1)^m\Big(1+\frac{a^2}{m^2{\pi}^2}\Big)\Big(\sin{\frac{bm{\pi}}{a}}\cos{\frac{cm{\pi}}{a}}+\sin{\frac{cm{\pi}}{a}}\cos{\frac{bm{\pi}}{a}}\Big)\\
  &+2\Big(1-\frac{a^2}{m^2{\pi}^2}\Big)\Big(\sin{\frac{bm{\pi}}{a}}\cos{\theta_2}+\sin{\frac{cm{\pi}}{a}}\cos{\theta_1}\Big),\\
    B=&\;a(-1)^m\Big(3+\frac{a^4}{m^4{\pi}^4}\Big)\sin{\frac{bm{\pi}}{a}}\sin{\frac{cm{\pi}}{a}}\\
    &+2\Big(1-\frac{a^2}{m^2{\pi}^2}\Big)\Big(a(-1)^m\cos(\theta_1-\theta_2)+b\cos{\frac{bm{\pi}}{a}}\cos{\theta_2}+c\cos{\frac{cm{\pi}}{a}}\cos{\theta_1}\Big)\\
    &+4\frac{a^3}{m^3{\pi}^3}(\sin{\frac{bm{\pi}}{a}}\cos{\theta_2}+\sin{\frac{cm{\pi}}{a}}\cos{\theta_1})\\
    &+4(-1)^m\frac{a^3}{m^3{\pi}^3}\Big(\sin{\frac{bm{\pi}}{a}}\cos{\frac{cm{\pi}}{a}}+\sin{\frac{cm{\pi}}{a}}\cos{\frac{bm{\pi}}{a}}\Big)\\
    &-2(-1)^m\Big(1+\frac{a^2}{m^2{\pi}^2}\Big)(a+b+c)\cos{\frac{bm{\pi}}{a}}\cos{\frac{cm{\pi}}{a}}\\
    &+2(-1)^m(b+c)\sin{\frac{bm{\pi}}{a}}\sin{\frac{cm{\pi}}{a}},\\
    A=&\;ac(-1)^m\Big(3+\frac{a^4}{m^4{\pi}^4}\Big)\sin{\frac{bm{\pi}}{a}}\cos{\frac{cm{\pi}}{a}}\\
    &+ab(-1)^m\Big(3+\frac{a^4}{m^4{\pi}^4}\Big)\cos{\frac{bm{\pi}}{a}}\sin{\frac{cm{\pi}}{a}}\\
    &-4a(-1)^m\frac{a^5}{m^5{\pi}^5}\sin{\frac{bm{\pi}}{a}}\sin{\frac{cm{\pi}}{a}}\\
     &+4\frac{a^3}{m^3{\pi}^3}\Big(a(-1)^m\cos(\theta_1-\theta_2)+b\cos{\frac{bm{\pi}}{a}}\cos{\theta_2}+c\cos{\frac{cm{\pi}}{a}}\cos{\theta_1}\Big)\\
      &-6\frac{a^4}{m^4{\pi}^4}(\sin{\frac{bm{\pi}}{a}}\cos{\theta_2}+\sin{\frac{cm{\pi}}{a}}\cos{\theta_1})\\
    &-\Big(1-\frac{a^2}{m^2{\pi}^2}\Big)\Big(b^2\sin{\frac{bm{\pi}}{a}}\cos{\theta_2}+c^2\sin{\frac{cm{\pi}}{a}}\cos{\theta_1}\Big)\\
    &-6(-1)^m\frac{a^4}{m^4{\pi}^4}\Big(\sin{\frac{bm{\pi}}{a}}\cos{\frac{cm{\pi}}{a}}+\sin{\frac{cm{\pi}}{a}}\cos{\frac{bm{\pi}}{a}}\Big)\\
    &+4(-1)^m\frac{a^3}{m^3{\pi}^3}\Big[(a+b+c)\cos{\frac{bm{\pi}}{a}}\cos{\frac{cm{\pi}}{a}}-(b+c)\sin{\frac{cm{\pi}}{a}}\sin{\frac{bm{\pi}}{a}}\Big]\\
    &+(-1)^m\Big(1+\frac{a^2}{m^2{\pi}^2}\Big)\Big[(a^2+b^2+c^2+2ab+2bc)\sin{\frac{bm{\pi}}{a}}\cos{\frac{cm{\pi}}{a}}\\
    &+(a^2+b^2+c^2+2bc+2ac)\cos{\frac{bm{\pi}}{a}}\sin{\frac{cm{\pi}}{a}}\Big].\\
 \end{aligned}
\end{equation*}
 %----------------%
In the leading order, we then have
 %----------------%
 \begin{equation*}
 \delta \approx -\frac{C}{B}.
 \end{equation*}
 %----------------%}
The $\mathcal{O}(m^0) = \mathcal{O}(1)$ terms in the numerator $C$ can be canceled out with the appropriate choice of $\theta_1$ and $\theta_2$ in accordance with the high-energy limit presented earlier, specifically  $\cos{\theta_2} = (-1)^m\cos{\frac{cm{\pi}}{a}}$ and $\cos{\theta_1} = (-1)^m\cos{\frac{bm{\pi}}{a}}$. At the same time, the $\mathcal{O}(m^{-2})$ terms present there remain and the expression becomes $C = -4(-1)^m\frac{a^2}{m^2{\pi}^2}\Big(\sin{\frac{bm{\pi}}{a}}\cos{\frac{cm{\pi}}{a}}+\sin{\frac{cm{\pi}}{a}}\cos{\frac{bm{\pi}}{a}}\Big)$, while the $\mathcal{O}(1)$ terms in the denominator $B$ are not influenced by such a choice, and, after some easy algebraic manipulations, we find $B = (-1)^m(5a+2b+2c)\sin{\frac{cm{\pi}}{a}}\sin{\frac{bm{\pi}}{a}} + \mathcal{O}(m^{-2})$. In the described situation, there are therefore gaps around points $k = \frac{m{\pi}}{a}$ of the halfwidth
 %----------------%
\begin{equation*}
    \delta =\frac{1}{m^2{\pi}^2}\frac{4a^2\Big(\sin{\frac{bm{\pi}}{a}}\cos{\frac{cm{\pi}}{a}}+\sin{\frac{cm{\pi}}{a}}\cos{\frac{bm{\pi}}{a}}\Big)}{(5a+2b+2c)\sin{\frac{cm{\pi}}{a}}\sin{\frac{bm{\pi}}{a}}}+\mathcal{O}(m^{-4}),
\end{equation*}
 %----------------%
which behave as $m^{-2}$ at the momentum scale as $m \to \infty$. Since the spectral condition \eqref{Dil_hex_spec_cond} is symmetric with respect to exchanges of any two lengths $a,\,b,\,c$, the same can be said also about points $k = \frac{m{\pi}}{b}$ and $k = \frac{m{\pi}}{c}$ (with an appropriate permutation of the lengths in the expression of $\delta$), all of which, together with $k = \frac{m{\pi}}{a}$, appear in the spectrum periodically with respect to momentum.

Of course, the situation would not be exactly the same when some or all of the lengths are commensurate, but it is similar. For definiteness, let us assume one commensurate pair only, e.g., $b$ and $m$ such that $\sin{m\pi} = \sin{\frac{bm{\pi}}{a}} = 0$. Due to the high energy limit, $\cos{\theta_1}$ must be chosen in the same way as before, being equal to $(-1)^m\cos{\frac{bm{\pi}}{a}} = (-1)^{m(b/a+1)}$, while $\cos{\theta_2}$ is yet free of such constraints. Then
 %----------------%
  \begin{equation*}
 \begin{aligned}
 C =&-4(-1)^{m(b/a+1)}\frac{a^2}{m^2{\pi}^2}\sin{\frac{cm{\pi}}{a}},\\
B=&\;2(-1)^{mb/a}(a+b)\cos{\theta_2}-2(-1)^m(-1)^{mb/a}(a+b)\cos{\frac{cm{\pi}}{a}}+ \mathcal{O}(m^{-2}),
 \end{aligned}
\end{equation*}
 %----------------%
 and we would get the same behavior, $\delta \approx m^{-2}$, unless $\cos{\theta_2}$ coincides with $(-1)^m\cos{\frac{cm{\pi}}{a}}$ (as in the high-energy limit in the incommensurate situation). In that case, we must include in our calculation also the $A$ term,
  %----------------%
  \begin{equation*}
 \begin{aligned}
 A =(-1)^{m(b/a+1)}(3ab+a^2+b^2+2bc+2ac)\sin{\frac{cm{\pi}}{a}}+\mathcal{O}(m^{-2})
 \end{aligned}
\end{equation*}
 %----------------%
and to express $\delta$ as a pair of solutions of a quadratic equation. We are interested in the asymptotic behavior only, which allows us to use the approximation
   %----------------%
  \begin{equation*}
 \begin{aligned}
\delta = -\frac{B}{2A} \pm \sqrt{\frac{B^2}{4A^2}-\frac{C}{A}}\;\; \Longrightarrow\;\; \delta \approx -\frac{B}{2A} \pm \sqrt{-\frac{C}{A}}\Big(1-\frac{B^2}{8AC}\Big)\;\;\; \textit{if}\;\; B^2 \ll AC.
 \end{aligned}
\end{equation*}
 %----------------%
Since $C = \mathcal{O}(m^{-2})$, $B = \mathcal{O}(m^{-2})$ and $A = \mathcal{O}(1)$, we are left with
   %----------------%
  \begin{equation*}
 \begin{aligned}
\delta = \pm \sqrt{-\frac{C}{A}} +\mathcal{O}(m^{-2}) = \pm \frac{1}{m{\pi}}\frac{2a}{\sqrt{3ab+a^2+b^2+2bc+2ac}} +\mathcal{O}(m^{-2})
 \end{aligned}
\end{equation*}
 %----------------%
 independently of the value of $\sin{\frac{cm{\pi}}{a}}$, assuming it is non-zero. Should it also be zero, then we have an example of a flat band - if $\sin{m\pi} = \sin{\frac{bm{\pi}}{a}} = \sin{\frac{cm{\pi}}{a}} = 0$, the whole spectral condition \eqref{Dil_hex_spec_cond} vanishes independently of $\theta_1$ and $\theta_2$. The argument regarding the permutation of lengths still holds.

 Let us now briefly mention two situations not included in the analysis above: the commensurability ratio in which $\sin{m\pi} = \cos{\frac{bm{\pi}}{a}} = 0$, and the case in which $\sin{m\pi} = \cos{\frac{bm{\pi}}{a}} = \cos{\frac{cm{\pi}}{a}} = 0$. In the first one, we end up with
  %----------------%
  \begin{equation*}
 \begin{aligned}
 C =&-4(-1)^{m(b/a+1)-1/2}\frac{a^2}{m^2{\pi}^2}\cos{\frac{cm{\pi}}{a}},\\
B=&(-1)^{m(b/a+1)-1/2}(5a+2b+2c)\sin{\frac{cm{\pi}}{a}} + \mathcal{O}(m^{-2}),
 \end{aligned}
\end{equation*}
 %----------------%
 and
  %----------------%
\begin{equation*}
    \delta =\frac{1}{m^2{\pi}^2}\frac{4a^2}{(5a+2b+2c)}\cot{\frac{cm{\pi}}{a}}+\mathcal{O}(m^{-4}),
\end{equation*}
 %----------------%
 which can be viewed as a limit case of incommensurate lengths. The second one once again leads to a point $\frac{m\pi}{a}$ being a part of the spectrum (not a flat band though), because then the choice $\cos{\theta_1} = \cos{\theta_2} = 0$ solves the spectral condition \eqref{Dil_hex_spec_cond}. To summarize the discussion:
\begin{itemize}
    \item There is an infinite number of spectral gaps for any possible combination of dilated hexagon lengths. Their widths are generally of the order $\mathcal{O}(m^{-2})$ at the momentum scale as $m\to\infty$ for incommensurate lengths, and they may be of order $\mathcal{O}(m^{-1})$ around some points if the lengths are commensurate; recall that all the gaps behave like that if the hexagon is equilateral.
\end{itemize}

 %%%%%%%%%%%%%%%%%%%%%%%%%%%%%%%%%
 \subsection{The low-energy limit}

 For $k \to 0$, we expand condition \eqref{Dil_hex_spec_cond} around the point $k=0$, getting
  %----------------%
  \begin{equation}\label{Dil_hex_spec_cond_small_k}
	k^3(-abc + 2(a+b+c) + 2(a\cos({\theta_2-\theta_1}) + b\cos{\theta_2} + c\cos{\theta_1})) = \mathcal{O}(k^{5}).
\end{equation}
 %----------------%
To decide whether there is an open gap around $k=0$ we have to find extrema of the left-hand side of \eqref{Dil_hex_spec_cond_small_k}. Finding the maximum is easy; since the edge lengths are positive, it is sufficient to choose $\theta_1 = \theta_2 = 0$. The minimum requires a little more care.   We reformulate Lemma 3.3 of \cite{ET15} denoting for a moment $a\cos({\theta_2-\theta_1}) + b\cos{\theta_2} + c\cos{\theta_1} =: f(\theta_1, \theta_2)$. If
  %----------------%
  \begin{equation*}
	\frac{1}{a}+\frac{1}{b}+\frac{1}{c}\le 2\max \Big\{\frac{1}{a},\frac{1}{b},\frac{1}{c}\Big\},
\end{equation*}
 %----------------%
then $\min_{\theta_1, \theta_2} f(\theta_1, \theta_2) = -a-b-c+2\min\{a,b,c\}$ and a spectral band in the vicinity of $k = 0$ exists provided
  %----------------%
  \begin{equation}\label{Dil_hex_band_cond_small_k_1}
	4\min\{a,b,c\} < abc < 4(a+b+c).
\end{equation}
 %----------------%
On the other hand, if
 %----------------%
  \begin{equation*}
	\frac{1}{a}+\frac{1}{b}+\frac{1}{c}\ge 2\max \Big\{\frac{1}{a},\frac{1}{b},\frac{1}{c}\Big\},
\end{equation*}
 %----------------%
 then $\min_{\theta_1, \theta_2} f(\theta_1, \theta_2) = -\frac{abc}{2}\big(\frac{1}{a^2}+\frac{1}{b^2}+\frac{1}{c^2}\big)$ and the band condition reads
   %----------------%
  \begin{equation}\label{Dil_hex_band_cond_small_k_2}
	2(a+b+c) - abc\Big(\frac{1}{a^2}+\frac{1}{b^2}+\frac{1}{c^2}\Big) < abc < 4(a+b+c).
\end{equation}
 %----------------%
If these conditions are, and each particular situation, satisfied, the positive spectrum extends to zero.

%%%%%%%%%%%%%%%%%%%%%%%%%%%%%%
\subsection{Negative spectrum}

As in the regular case, the respective spectral condition is obtained directly from \eqref{Dil_hex_spec_cond_better} through replacing real $k$ with $k = i\kappa$. Referring to Theorem 2.6 of \cite{BET22}, we know that there are at most two negative spectral bands, as at each of the two vertices in the elementary cell the matrix $-R$ has exactly one eigenvalue in the upper complex plane. Moreover, the above expansion of the spectral condition in the leading order is the same as for the positive spectrum. The higher negative spectral band then extends to zero only if it is true for the lowest positive one, otherwise the spectrum has a gap around $k = 0$.

%%%%%%%%%%%%%%%%%%%%%%%%%%%%%%%%%%%%%%%%%
\subsection*{Data availability statement}

Data are available in the article.

%%%%%%%%%%%%%%%%%%%%%%%%%%%%%%%%%%%%%%%%%
\subsection*{Conflict of interest}

The authors have no conflict of interest.

%%%%%%%%%%%%%%%%%%%%%%%%%%%%%%%%%%%%%%%%%
\section*{Acknowledgements}

The research was supported by the European Union's Horizon 2020 research and innovation programme under the Marie Sklodowska-Curie grant agreement No 873071.

%%%%%%%%%%%%%%%%%%%%%%%%%%%%%%%%%%%%%%%%%

%\newpage
%\def\cprime{$'$}

\end{document}